\documentclass[12pt]{article}

\usepackage{hyperref}
\def\be{\begin{equation}}
\def\ee{\end{equation}}
\def\ba{\begin{array}}
\def\ea{\end{array}}
\def\d{\partial}

\def\ba{\begin{array}}
\def\ea{\end{array}}
\def\f{\frac}
\def\vac{|0\rangle}
\def\ie{{\it i.e.}~}
\def\eg{{\it e.g.}~}

\begin{document}

\begin{flushright}
Workshop on HS gauge theories\\
Bruxelles, May  2004\\
\end{flushright}
\vspace{0.51cm}

\begin{center}

{\small\ttfamily ROM2F/05/03 \ FIAN/TD/05/04\hspace*{0.8cm} }

\end{center}


\begin{center}

\renewcommand{\thefootnote}{\fnsymbol{footnote}}

{\bf\Large  ``Massive'' Higher Spin Multiplets \\
\bigskip
and Holography } \vspace{1cm}

\addtocounter{footnote}{1} \textbf{ M.~Bianchi}
\footnote{\texttt{Massimo.Bianchi@roma2.infn.it}} and
\textbf{V.~Didenko \footnote{\texttt{Didenko@lpi.ru}}}

\vspace{0.5cm} \textit{
Dipartimento di Fisica and Sezione INFN \\
Universit\`a di Roma ``Tor Vergata''\\
00133 Rome, Italy}\par

\vspace{0.3cm}

\textit{ I.E.Tamm Department of Theoretical Physics, \\
P.N.Lebedev Physical Institute,\\
Leninsky prospect 53, 119991, Moscow, Russia}\par

 \setcounter{footnote}{0}
\end{center}

\begin{abstract}
We review the extrapolation of the single-particle string spectrum on $AdS_5\times S^5$ to the Higher Spin enhancement point and the successful comparison of the resulting spectrum with the one of single-trace gauge-invariant operators in ${\cal N}=4$ supersymmetric Yang-Mills theory. We also describe how to decompose the common spectrum in terms of massless and massive
representations of the relevant Higher Spin symmetry group.

Based on the lecture delivered by M. Bianchi at the First Solvay  Conference
on Higher-Spin Gauge Theories held in Bruxelles, on May 12-14, 2004.
\end{abstract}
\vspace{1cm}

\section{Introduction}

We present an overview of the work done by one of the authors
(M.B.) in collaboration with N. Beisert, J.F. Morales and H.
Samtleben \cite{Bianchi:2003wx, Beisert:2003te,
Beisert:2004di}\footnote{A shorter account can be found in
\cite{MBstring}}. After giving some historical motivations for the
interest in higher spin (HS) gauge fields and currents, we very
briefly and schematically review some of the achievements of the
holographic $AdS/CFT$ correspondence\footnote{For recent reviews
see \eg \cite{Aharony:1999ti, D'Hoker:2002aw, Bianchi:2000vh,
Tseytlin:2003ii}.}. We mostly but not exclusively focus on
protected observables that do not change as we vary 't Hooft
coupling constant $\lambda = g^2_{_{YM}} N$. We then discuss how
the single-particle string spectrum on $AdS_5\times S^5$ can be
extrapolated to the HS enhancement point \cite{WittenJHS60,
Sundborg:1999ue,Polyakov:2001af,Sezgin:2001zs, Sezgin:2001yf,
Sezgin:2002rt} and how it can be successfully compared with the
spectrum of single-trace gauge-invariant operators in ${\cal N}=4$
supersymmetric Yang-Mills (SYM) theory
\cite{Bianchi:2003wx,Beisert:2003te}. To achieve the goal we rely
on the aid of Polya theory \cite{Polya}. We also decompose the
resulting spectrum in terms of massless and massive
representations of the relevant HS symmetry group
\cite{Beisert:2004di}. Eventually, we concentrate our attention on
the generalization of the HS current multiplets, \ie semishort
multiplets, which saturate a unitary bound at the HS enhancement
point and group into long ones as we turn on interactions.
Finally, draw our conclusions and perspectives for future work.
Properties of HS gauge theories are extensively covered by other
contributions to this conference
\cite{Bouatta:2004kk,Petkoubrux,Vasbrux,Sundbrux,Hullbrux} as well
as the reviews \eg \cite{Vasiliev:2004qz,Sorokin:2004ie,MBRTN}.

\section{Historical motivations}

The physical interest in HS currents dates back to the studies of
QCD processes, such as deep inelastic scattering, where the
structure of hadrons was probed by electrons or neutrinos. The
process is studied at a scale $Q^2=-q^2$, related to the momentum
$q$ transferred by the photon, which is much larger than the
typical mass parameter of the theory $\Lambda_{QCD}$.

 The fraction of momentum carried by the struck ``parton", \ie one of the
 hadron's constituents, is given by the Bjorken variable
\be 0\leq\xi=x_{B}=\f{Q^2}{2 P \cdot q}\leq 1\,, \ee
where $P$ is the momentum of the hadron. Note that $\xi$ is
kinematically fixed. The optical theorem relates the amplitude of
the process to the forward Compton amplitude $W^{\mu\nu}$
\[
W^{\mu\nu}=W_{1}(x_B)\Big(\f{q^{\mu}q^{\nu}}{q^2}-\eta^{\mu\nu}\Big)
+W_{2}(x_B)\Big(P^{\mu}-q^{\mu}\f{P\cdot
q}{q^2}\Big)\Big(P^{\nu}-q^{\nu}\f{P\cdot q}{q^2}\Big)=
\]
\be =i\int d^4x e^{iqx}\int_{0}^{1}\f{d\xi}{\xi}\sum_{i}f_{i}(\xi)
\langle q_{i}(\xi P)|T(J^{\mu}(x)J^{\nu}(0))|q_{i}(\xi P)\rangle\,,
\ee
where $W_{1}(x)$ and $W_{2}(x)$ are scalar structure functions and
$f_{i}(\xi)$ are the parton distribution functions, which depend
on non-perturbative dynamical effects such as  confinement.
Parity-violating terms which can appear in weak interactions are
omitted for simplicity. For non interacting spin $1/2$ partons,
the structure functions satisfy Callan-Gross relations
\be {\mathcal Im} W_{1}=\pi \sum_{i}e_{i}^2 f_{i}\,,\qquad  \qquad  {\mathcal Im}
W_{2}=\f{4 x_{B}}{Q^2}{\mathcal Im} W_{1}\,. \ee

Operator Product Expansion (OPE) yields \be W^{\mu\nu}=\sum_{i}
e_{i}^2\Big[ \sum_{M=0}^{\infty}\f
{P^{\mu}P^{\nu}}{Q^2}\Big(\f{2P\cdot
q}{Q^2}\Big)^{M-2}A_{i}^{(n)}(Q^2)-
\f{1}{4}\eta^{\mu\nu}\sum_{M=0}^{\infty}\Big(\f{2P\cdot
q}{Q^2}\Big)^{M} A_{i}^{(n)}(Q^2)\Big]+\dots\,, \ee

where the dominant contribution arises from operators with lowest
twist $\tau=\Delta-s = 2 + \dots$ The non-perturbative information
is coded in the coefficients $A_{i}^{(n)}$ which can be related to
the matrix elements in the hadronic state of totally symmetric and
traceless HS currents built out of  quark fields $\psi$

\be A_{i}^{(n)}:\qquad \langle P|\bar{\psi_{i}}\gamma^{(\mu_{1}}
D^{\mu_{2}}\dots D^{\mu_{n})}\psi_{i}|P\rangle \ee

Higher twist operators and flavour non-singlet operators, such as
\[
\eta_{\lambda_{i}\lambda_{j}}\dots\eta_{\tau_{\nu}\rho_{\tau}}\dots
\bar{\psi_{i}}D^{\mu_{1}}\dots D^{\mu_{k}}F^{\rho_{1}\nu_{1}}
D^{\lambda_{1}}\dots D^{\lambda_{\rho}} F^{\rho_{2}\nu_{2}}\dots
D^{\tau_{1}}\dots D^{\tau_{\nu}}\psi_{j}\,,
\]
can appear in non-diagonal OPE's and  produce mixing with purely
gluonic operators. Their contribution to the OPE of two currents
is suppressed at large $Q^2$.

\subsection{Broken scale invariance}

It is well known that QCD is only approximately scale invariant in
the far UV regime of very large $Q^2$ where it exposes asymptotic
freedom \cite{nobel}. Scale invariance is indeed broken by quantum
effects, such as vacuum polarization that yields $\beta\neq 0$,
and dimensional transmutation generates the QCD scale \be
\Lambda_{QCD} = \mu e^{-\f{8\pi^2}{b g^2(\mu)}}\,. \ee

The coefficient functions $A_{i}^{(n)} (Q^2)$ turn out to be
Mellin transforms of the parton distributions and evolve with the
scale $Q^2$ due to quantum effects. Depending on the parity of
$n$, the parton and antiparton distributions contribute with the
same or opposite sign. Defining $f^{\pm}_{i}=f_{i}\pm
\bar{f_{i}}$, one has \be A_{i}^{(n)} (Q^2) =
\int_{0}^{1}d\xi\xi^{n-1}f_{i}^{+}(\xi, Q^2) \ee in the case of
even $n$ and similarly \be A_{i}^{(n)}=\int_{0}^{1}d\xi
\xi^{n-1}f_{i}^{-}(\xi,Q^2)\,, \ee in the case of odd $n$.

Parton distributions satisfy sum rules arising from global
conservation laws stating \eg that the net numbers of constituents of a
given hadron do not depend on the scale \be \int_{0}^{1}d\xi
f_{i}^{-}(\xi, Q^2)=n_{i}\,, \ee where $n_{i}$ is independent
of $Q^2$. For instance we know that protons are made of two up
quarks $n_u=2$ and one down quark $n_d=1$ at each scale. Similarly
the total momentum, including the gluonic contribution labelled by $g$,
should be equal to the momentum of the hadron, and one obtains
another sum rule of the form
\be \sum_{i}\langle x_{i}\rangle+\langle x_{g}\rangle=1 \ee

The evolution of the parton distributions is governed by
Altarelli-Parisi (AP) equations \cite{Altarelli}\footnote{Closely
related equations were found by Gribov and Lipatov for QED
\cite{Grib}.}. For odd $n$, there is no operator mixing and the
GLAP equations are "diagonal"
\be \f{d}{dt}A^{(n)}_{i}(t)=\f{\alpha_{s}(t)}{2\pi}
\hat{P}^{(n)}A_{i}^{(n)}(t)\,,\quad t=\log{Q^2}\quad\,,  \ee
To lowest order, the relevant kernel is
 \be\label{altar} \hat{P}^{(n)}=\int_{0}^{1} dz
z^{n-1}P_{q\leftarrow q}(z)= \int_{0}^{1} dz
z^{n-1}\Big(\f{4}{3}\Big)
\Big[\f{1+z^2}{(1-z)_{+}}+\f{3}{2}\delta(1-z)\Big]\,. \ee 
that can be identified with the Mellin
transform of the probability for a spin 1/2 parton to emit an
almost collinear gluon. The result turns out to be simply given by
\be
\hat{P}^{(n)} = -\f{2}{3}\Big[1+4\sum_{k=2}^{n}\f{1}{k}-\f{2}{n(n+1)}\Big]
\quad . \ee The presence of the harmonic numbers calls for a
deeper, possibly number theoretic, interpretation and implies that
the anomalous dimensions $\gamma_{S}$ of HS currents at one loop
behave as $\gamma_{S}\sim \log{S}$ for $S\gg 1$. Remarkably
enough, the same leading behaviour holds true at two and higher
loops \cite{Petronzio}.

For even $n$ there is mixing with purely gluonic HS currents of
the form \be J_{v}^{(\mu_{1}\dots
\mu_{n})|}=F_{\lambda}^{(\mu_{1}}D^{\mu_{2}}\dots
D^{\mu_{n-1}}F^{\mu_{n})|\lambda}\quad , \ee In the holographic
perspective twist two HS currents should correspond to nearly
massless HS gauge fields in the bulk theory. Moreover, in ${\cal
N}=4$ SYM one has to take into account twist two currents that are
made of scalars and derivatives thereof \be
J_{\phi}^{(\mu_{1}\dots \mu_{n})|}=\varphi_{i}D^{\mu_{1}}\dots
D^{\mu_{n}}\varphi^{i} \quad . \ee Although ${\cal N}=4$ SYM
theory is an exact superconformal field theory (SCFT) even at the
quantum level, thanks to the absence of UV divergences  that
guarantees the vanishing of the $\beta$-function, composite
operators can have nonvanishing anomalous dimensions.

\subsection{Anomalous dimensions}

We now turn to discuss anomalous dimensions and  unitary bounds.
In a CFT$_D$ a spin $S$ current with scaling dimension
$\Delta=S+D-2$ is necessarily conserved. For instance, a vector
current with $S=1$ and $\Delta=3$ in $D=4$ has a unique conformal
invariant 2-point function of the form \be \langle
J^{\mu}(x)J^{\nu}(0)\rangle=(\d^{\mu}\d^{\nu}-\d^2\delta^{\mu\nu})
\f{1}{x^4}\,,\qquad (D=4)\quad , \ee that implies its
conservation. For $\Delta = 3 + \gamma$ one finds instead \be
\langle \hat{J}^{\mu}(x)\hat{J}^{\nu}(0)\rangle=
\f{1}{x^{2\gamma}} (\d^{\mu}\d^{\nu}-\d^2\delta^{\mu\nu})
\f{1}{x^4} \quad , \ee that leads to the (anomalous) violation of
the current. Anomalous dimensions of HS currents satisfy
positivity constraints. For instance, in $D=4$, a scaling operator
carrying non-vanishing Lorentz spins $j_{L}$ and $j_{R}$ satisfies
a unitary bound of the form \be\label{bound} \Delta\geq
2+j_{L}+j_{R} \ee At the threshold null states of the form
$A=\d^{\mu}J_{\mu}$ (dis)appear. When $j_{L}=0$ or $j_{R}=0$, \eg
for spin 1/2 fermions and scalars, the unitary bound (\ref{bound})
takes a slightly different form \be \Delta\geq 1+j \quad . \ee The
identity is the only (trivial) operator with vanishing scaling
dimension.

In ${\cal N}=4$ SYM the situation gets a little bit more involved
\cite{DobrevPetkova, Dolan:2002zh, Heslop:2003xu,
Andrianopoli:1998ut}.  The highest weight state (HWS) of a unitary
irreducible representation (UIR) of $(P)SU(2,2|4)\subset U(2,2|4)
= SU(2,2|4)\times U(1)_{B}$ can be labelled by

\be \mathcal{D}\Big(\Delta,(j_{L},j_{R}), [q_{1},p,q_{2}]; B, C;
L, P\Big)\,, \ee

where $\Delta$ is the dimension, $(j_{L},j_{R})$ are the Lorentz
spins, $[q_{1}, p, q_{2}]$ are the Dynkin labels of an $SU(4)$
R-symmetry representation. The central charge $C$, which commutes
with all the remaining generators but can appear in the
anticommutator of the supercharges, and the ``bonus" $U(1)_B$
charge, related to an external automorphism of $SU(2,2|4)$, play a
subtle role in the HS generalization of the superconformal group.
The discrete quantum number $P$ can be associated with the
transposition of the gauge group generators or with a generalized
world-sheet parity of the type IIB superstring. Finally the length
$L$ of an operator or a string state which is related to twist,
but does not exactly coincide with it, is a good quantum number up
to order one loop in $\lambda$.

Setting $C=0$ and neglecting the  $U(1)_B$ charge, there are three
types of UIR representations of $PSU(2,2|4)$ relevant for
the description of ${\cal N}=4$ SYM theory.

\begin{itemize}

\item type A

For generic $(j_{L},j_{R})$ and $[q_{1},p,q_{2}]$, one has
 \be
 \Delta\geq 2+j_{L}+j_{R}+q_{1}+p+q_{2}\quad ,
 \ee
that generalizes the unitary bound of the conformal group. The
bound is saturated by the HWS's of ``semishort" multiplets of
several different kinds. Current-type multiplets correspond to
$j_{L}=j_{R}=S/2$ and $p=q_{1}=q_{2}=0$.  Kaluza-Klein (KK)
excitations of order $p$ to $j_{L}=j_{R}=0$ and $q_{1}=q_{2}=0$.
Above the bound, multiplets are long and comprise $2^16$ components times the dimension of the HWS.

\item type B

For $j_{L}j_{R}=0$, say $j_{L}=0$ and $j_{R}=j$, one has
 \be
 \Delta\geq 1+j_{R} + {1\over 2} q_1 + p \quad ,
 \ee
that generalizes the unitary bound  of the conformal group. At
threshold one finds 1/8 BPS multiplets.

\item type C

For $j_{L}=j_{R}=0$ and $q_{1}=q_{2}=q$, one has
\be
 \Delta = 2q + p \quad   ,
 \ee
the resulting UIR is 1/4 BPS if $q\neq 0$ and  1/2 BPS when $q=0$.
In the 1/2 BPS  case, the number of components is
$2^8p^2(p^2-1)/12$, the multiplet is protected  against quantum
corrections and is ultrashort for $p\le 3$. For $p=1$ one has the
singleton, with 8 bosonic and as many fermionic components, that
corresponds to the elementary (abelian) vector multiplet. In the
1/4 BPS case, if the HWS remains a primary when interactions are
turned on, the multiplet remains 1/4 BPS and short and protected
against quantum corrections. For single trace operators, however,
the HWS's all become  superdescendants and acquire anomalous
dimensions in a pantagruelic Higgs-like mechanism that deserves to
be called {\it ``La Grande Bouffe"}.

\end{itemize}

\section{Lessons from $AdS/CFT$}

Before entering the main part of the lecture,  it may be useful to
summarize what we have learned from the holographic $AdS/CFT$
correspondence \cite{Aharony:1999ti, D'Hoker:2002aw,
Bianchi:2000vh, Tseytlin:2003ii}. Let us list some of the
important lessons.

\begin{itemize}

\item The spectrum of $1/2$ BPS single-trace gauge-invariant
operators at large $N$ matches perfectly  well with the
Kaluza-Klein spectrum of type  IIB supergravity on $S^5$.

\item The 3-point functions of chiral primary  operators (CPO's)
${\cal Q}_{p}$, which are HWS's of 1/2 BPS multiplets, \be \langle
{\cal Q}_{p_1}(x_{1}){\cal Q}_{p_2}(x_2){\cal Q}_{p_3}(x_3)\rangle
= C(p_{1}, p_{2}, p_{3};N)
\prod_{i<j}x_{ij}^{-2(l_i+l_j-\Sigma)}\,, \ee are not renormalized
by interactions and, as shown in \cite{Seiberg}, only depend on
the quantum numbers $p_{i}$, associated with the spherical
harmonics on $S^5$, and on the number of colors $N$, but neither
on the gauge coupling $g_{YM}$ nor on the vacuum angle
$\vartheta_{YM}$ \cite{dinst}.

\item There are some additional observables that are not
renormalized. In particular, extremal and next-to extremal
$n$-point correlators of CPO's are exactly the same as in the free
theory \cite{extrem}. A  correlator of CPO's is (next-to) extremal
when $p_0$ is the sum (minus two) of the remaining $p_i$ and one
finds \be \langle Q_{l_0}(x_0)Q_{l_1}(x_1)\dots
Q_{l_n}(x_n)\rangle = G_{n+1}^{free}\,,\qquad
l_{0}=\sum_{i=1}^{n}l_{i}(-2)\,. \ee

\item For near extremal correlators   with
$l_0=\sum_{i}l_{i}(-4,-6,\dots)$ one has partial
non-renormalization \cite{partial}. In practise these correlation
functions depend on lesser structures than naively expected in
generic conformal field theory. Nevertheless these results are
consequences of $PSU(2,2|4)$ invariance.

\item Instanton effects ${\cal N}=4$ SYM  correspond to
$D$-instanton effects in type IIB superstring. In particular,
certain higher derivative terms in the superstring effective
action on $AdS_5\times S^5$ are exactly reproduced by instanton
dominated correlators on the boundary.

\item (Partial) non-renormalization of "BPS" Wilson  loops holds.
For instance two parallel lines do not receive quantum corrections
\cite{maldawilson} while circular loops receive perturbative
contributions only from rainbow diagrams \cite{rainbow} and
non-perturbative contributions from instantons \cite{bgkwilson}.

\item The RG flows induced by  deformations of the boundary CFT
are holographically described by domain wall solutions in the
bulk. In particular, it is possible to prove the holographic
$c$-theorem. Indeed one can build a holographic $c$-function
\cite{holocthe} \be
\beta^{i}=\dot{\phi}^{i}=\f{\phi'(r)}{A'(r)}\,,\qquad \dot
c_{H}=-G_{ij}\beta^{i}\beta^{j}\leq 0\, , \ee and prove that it be
monotonically decreasing along the flow.

\item There is a nice way to reproduce anomaly  in ${\cal N}=4$
arising upon coupling the theory to external gravity or other
backgrounds. In particular the holographic trace anomaly reads
\cite{hensken} \be \langle {\cal T}_{\mu}^{\mu}\rangle_{(g_{\mu\nu})}=
\f{N^{2}}{4}(R_{\mu\nu}R^{\mu\nu}-\f{1}{8}R^2)\, . \ee Quite
remarkably, the structure of the anomaly implies \be c_{H}=a_{H}
\ee at least at large $N$. This is simple and powerful constraint
on the CFT's that admit a holographic dual description. The
techniques of holographic renormalization \cite{Bianchi:2001de,
Bianchi:2001kw,holoren} have been developed to the point that one
can reliably compute not only the spectrum of superglueball states
\cite{superglu} but also three-point amplitudes and the associated
decay rates \cite{3point}.

\item Very encouraging results come from the recent work on string
solitons with large spin or large  charges \cite{Tseytlin:2003ii},
which qualitatively reproduce gauge theory expectations. In
particular, the scaling 
\be
\sum_{k=1}^{S}\f{1}{k}\sim \log S\, \ee for long strings with large spin
$S$ on $AdS_{5}\times S^5$ has been found \cite{Tseytlin:2003ii}.
Moreover, the BMN limit \cite{Berenstein:2002jq}, describing
operators with large $R$ charge, is believed to be dual to string
theory on a pp-wave background emerging from the Penrose limit of
$AdS_{5}\times S^5$. For BMN operators, with a small number of
impurities, light-cone quantization of the superstring suggests a
close-form expression for the dimension as a function of coupling
$\lambda$. For two-impurity BMN operators one has
 \be
\Delta=J+\sum_{n}N_{n}\sqrt{1+\f{\lambda n^2}{J^2}}\,.
\ee where $J$ is the R-charge.
\end{itemize}

\section{Stringy $AdS_{5}\times S^{5}$ and higher spin holography}

In the boundary CFT the HS symmetry enhancement point is at
$\lambda=0$, so one may naively expect it to correspond to zero
radius for $AdS_{5}\times S^{5}$. Actually, there might be
corrections to $R^2=\alpha' \sqrt{\lambda}$ for $\lambda\ll 1$ and
we would argue that it is not unreasonable to expect the
higher-spin enhancement point to coincide with the self-dual point
$R^2\sim \alpha'$. Ideally, one would like to determine the string
spectrum by (covariant) quantization in $AdS_{5}\times S^{5}$
background. However the presence of a R-R background has prevented
a satisfactory resolution of the problem so far despite some
progress in this direction \cite{Berkovits:2002zk}.  Since we are
far from a full understanding of stringy effects at small radius
we have to devise an alternative strategy.

In \cite{Bianchi:2003wx} we computed the KK spectrum by naive
dimensional reduction on the sphere and then  extrapolated it to
small radius, \ie to the HS symmetry enhancement point. As we
momentarily see, group theory techniques essentially determine all
the quantum numbers, except for the scaling dimension, dual to the
AdS mass. In order to produce a formula valid for all states at
the HS point, we first exploit HS symmetry and derive a formula
for the dimension of the massless HS fields. Then we take the BMN
limit \cite{Berenstein:2002jq} as a hint and extend the formula
so as to encompass the full spectrum. The final simple and
effective formula does not only reproduce the HS massless
multiplets as well as their KK excitations but does also describe
genuinely massive states, which are always part of long
multiplets. Finally, we compare the resulting string spectrum at
the HS enhancement point in the bulk with the spectrum of free
${\cal N}=4$ SYM theory on the AdS boundary. Clearly the matching
of the spectrum is a sign that we are on the right track, but it
is by no means a rigorous proof.

In order to study the string spectrum on $AdS_{5}\times S^{5}$, we
started with the GS formalism and built the spectrum of type IIB
superstrings in the light-cone gauge in flat ten-dimensional
space-time. The little group is $SO(8)$ for massless states and
$SO(9)$ for massive ones. The chiral worldsheet supermultiplets
are described by \be Q_{s}=8_v-8_s\,,\qquad Q_c=8_v-8_c\,, \ee
(see \cite{Bianchi:2003wx} for notations and details). At level 0
one has the massless supergravity multiplet \be l=0\qquad Q_s
Q_s=T_0\qquad \textnormal{(supergravity: $128_B-128_F$)} \ee At
level 1 there are $2^{16}$ states as a result of the enhancement
of $SO(8)$ to $SO(9)$ (128-fermions, 84-totally symmetric tensors
and 44-antisymmetric) \be l=1\qquad
Q_{s}^2Q_{c}^2=T_1\qquad\textnormal{($2^{16}$ states:
$(44+84-128)^2$)} \ee

Similarly at level l=2
\be\label{nine}
l=2\qquad Q_{c}^2(Q_s+Q_s\cdot Q_s)=T_1\times (1+8_v)=T_1\times
(9)
\ee
and so on
\be
l\qquad \dots=T_1\times v_l^2\qquad (v_1=1, v_2=9,\dots)\,.
\ee

The important thing is that after building the spectrum one has to
rewrite each level of the spectrum in terms of $T_1$, comprising
$2^{16}$ states. Eventually, $T_1$ turns out to provide us with a
representation of the superconformal group. What  remains per each
chirality will be called $v_{l}$. Combining with the opposite
world-sheet chirality one gets  $v_{l}^2$ as ground states.

In order to extend the analysis to $AdS_{5}\times S^{5}$, \ie
perform the naive KK reduction, requires identifying which kinds
of representations of the $S^{5}$ isometry group $SO(6)$ appear
associated to a given representation of $SO(5)$. The latter arises
from the decomposition $SO(9)\rightarrow SO(4)\times SO(5)$ for
the massive states in flat space-time. Group theory yields the
answer: only those representations of $SO(6)$ appear in the
spectrum which contain the given representation of $SO(5)$ under
$SO(6)\rightarrow SO(5)$.

Thus, after diagonalizing the wave equation for the bulk fields
\[
\Phi(x,y)=\sum X_{AdS}(x)Y_{S^5}(y)
\]
the spectrum of a string on $AdS_{5}\times S^{5}$ assembles into
representations of $SU(4)\approx SO(6)$,  which are essentially
given by spherical harmonics, with AdS mass \be
R^2M^{2}_{\Phi}=\Delta(\Delta-4)-\Delta_{min}(\Delta_{min}-4)
\leftrightarrow C_{2}[SU(2,2|4)]\,. \ee

More explicitly, the wave equation can be written as
\be
(\nabla^{2}_{AdS_{5}\times S^5}-M_{\Phi}^2)\Phi_{\{\mu\}\{i\}}=0
\,,\qquad \{\mu\}\in R_{SO(4,1)}\,,\quad \{i\}\in R_{SO(5)}
\ee
and one gets
\be
\Phi_{\{\mu\}\{i\}}= \sum_{[kpq]} X_{\{\mu\}}^{[kpq]}(x)
Y_{\{i\}}^{[kpq]}(y)\,,
\ee
where $[kpq]\in SO(6)$, the isometry group of $S^5$ and
\be
\nabla^{2}_{S^5}Y_{\{i\}}^{[kpq]}=-\f{1}{R^2}\Big(
C_{2}[SO(6)]-C_2[SO(5)]\Big)Y_{\{i\}}^{[kpq]}\,.
\ee

The KK tower build on the top of $SO(5)$ representation is given the
following direct sum of $SO(6)$ representation
\be
KK_{[mn]}=\sum_{r=0}^{m}\sum_{s=0}^{n}\sum_{p=m-r}^{\infty} [r+s;
p; r+n-s]+\sum_{r=0}^{m-1}\sum_{s=0}^{n-1}\sum_{p=m-r-1}^{\infty}
[r+s+1; p; r+n-s]\,,
\ee

where $[mn]$ are the two Dynkin labels of an $SO(5)$-representation.

The remaining $SO(4,2)$ quantum numbers, required to perform a
lift to HWS representations ${\cal D}$  of $PSU(2,2|4)\supset
SO(4,2)\times SO(6)\supset SO(4)\times SO(5)$, are the Lorentz
spins $j_{L}$ and $j_{R}$, and the scaling dimension $\Delta$ \be
D\{\Delta; (j_L, j_R); [k,p,q]\}= (1+Q+\dots +Q^{16})\Psi^{(\Delta,
j_{L}, j_{R})}_{[k,p,q]}\,. \ee

For instance, at level $l=1$, which corresponds to ${\cal D}\{2, (00);
[000]\}\equiv \hat{T}_{1}^{(2)}$, the spectrum of KK excitations
assembles into an infinite number of $SU(2,2|4)$ representations
\be H_{1}^{KK}=\sum_{M=0}^{\infty} [0n0]^{M}_{(0,0)}
\hat{T}_{1}^{(2)} = \sum_{M=0}^{\infty}{\cal D}\{\Delta_{0}=2+n, (00),
[0n0]\}\,. \ee The HWS's in this formula have dimensions $\Delta_0
= 2+n$, spin-0 and belong to $SU(4)$ representation $[0n0]$, which
describes exactly the spherical harmonics. As we have already
mentioned, so far the dimension $\Delta_0$, at the HS point,  is
postulated so as to get the correct massless HS fields in the
bulk.

For $l=2$ the situation is slightly more involved, because
{\bf{9}} in (\ref{nine}) in $v_l$ is neither a representation of
the $SO(9,1)$ nor a representation of the $SO(4,2)\times SO(6)$.
In this case, the correct way to proceed is to first  decompose
${\bf{9}}\rightarrow{\bf{10}}-{\bf{1}}$ and then
${\bf{10}}\rightarrow{\bf{6}}+{\bf{4}}$ \be
\underline{9}\rightarrow [010]^{1}_{(00)}+[000]^{1}_{(\f{1}{2},
\f{1}{2})}-[000]_{(00)}^2\sim\underline{10}-\underline{1}\,. \ee
The corresponding KK-tower has a form
\[
 H_{2}^{KK}=\sum_{M=0}^{\infty}[0n0]^{M}_{00}\times
 \hat{T}_{1}^{(2)}\times \{[020]^{2}_{(00)}+[101]^{2}_{(00)}+
 [000]^{2}_{(00)}+ 2[010]^{2}_{(\f{1}{2},\f{1}{2})}+
\]
\be
 +[000]^{2}_{(11)}+[000]^{2}_{(10)}+[000]_{(01)}^2+
[000]^{2}_{(00)}+[000]^{4}_{(00)}-2[010]^{3}_{(00)}-
2[000]^{3}_{(\f{1}{2}, \f{1}{2})}\}\,.
\ee

It is worth stressing that negative multiplicities cause no
problem as they cancel in infinite sum over $n$,  precisely when
the dimension is chosen properly.

The analysis of higher levels is analogous  though slightly more
involved. One has \be H_l=\sum_{0}^{\infty}[0n0]_{00}^{M}\times
\hat{T}_{l}^{(2)} \times (v_{l}^2)\,, \ee with the decomposition
\be v_{l}^2=[000]_{(l-1,l-1)}^{\Delta}+\dots\,. \ee

\subsection{Exploiting HS symmetry}

The superconformal group $PSU(2,2|4)$ admits a HS symmetry
extension, called  $HS(2,2|4)$ extension \cite{Vas, Sezgin:2001zs,
Sezgin:2001yf, Sezgin:2002rt}. In \cite{Sezgin:2001zs,
Sezgin:2001yf, Sezgin:2002rt} Sezgin and Sundell have shown that
the superstring states belonging to the first Regge trajectory on
$AdS$ can be put in one to one correspondence with the physical
states in the master fields of Vasiliev's theory
\cite{Vasiliev:2004qz}. The HS fields which have maximum spin, \ie
$S_{max}=2l+2$ at level $l$, are dual to twist 2 currents, which
are conserved at vanishing coupling. Including their KK
excitations, one is lead to conjecture the following formula for
their scaling dimensions \be \Delta_{0}=2l+n \label{dimnaive} \ee
at the HS enhancement point. Now, at $\lambda\neq 0$, as we said
in the introduction, {\it "La Grande Bouffe"} happens, since HS
multiplets start to "eat" lower spin multiplets. For example, the
short and "massless" ${\cal N}=4$ Konishi multiplet combines with
three more multiplets \be
K_{\textnormal{long}}=K_{\textnormal{short}}+K_{\f{1}{8}}+
K^{*}_{\f{1}{8}}+K_{\f{1}{4}} \ee and becomes long and massive.
The classically conserved currents in the Konishi multiplet are
violated by the supersymmetric Konishi anomaly \be
\bar{D}^{A}_{\dot{\alpha}}\bar{D}^{B\dot{\alpha}} K = gTr(W^{AE}
[W^{BF},\bar{W}_{EF}])+g^2D^{\alpha}_{E}D_{\alpha F} Tr(W^{AE}
W^{BF})\,, \ee where $W^{AB}$ is the twisted chiral multiplet
describing the ${\cal N}=4$ singleton. In passing, the  anomalous
dimension of the Konishi multiplet is known up to three loops
\cite{Bianchi:2001cm, Bianchi2loop, Beisert:2004ry} \be
\gamma_{K}^{^{\rm 1-loop}}=3\f{g^2 N}{4\pi^2} \, , \quad
\gamma_{k}^{^{\rm 2-loop}}=-3\f{(g^2N)^2}{(4\pi^2)^2} \, ,
\quad \gamma_{k}^{^{\rm 3-loop}}=21\f{(g^2N)^4}{(4\pi^2)^4}
\quad, \ee whereas the anomalous dimensions of many other
multiplets were computed by using both old-fashioned
field-theoretical methods as well as modern and sophisticated
techniques based on the integrability of the  super-spin chain
capturing the dynamics of the ${\cal N}=4$ dilatation operator
\cite{Beisert:2004ry}.

A systematic comparison with the operator  spectrum of free ${\cal
N}=4$ SYM, may not forgo the knowledge of a mass formula
encompassing all string states at the HS enhancement. Remarkably
enough such a formula was ''derived" in \cite{Beisert:2003te}.
Consideration of the pp-wave limit of $AdS_5 \times S^5$ indeed
suggests the following formula \be\label{bmndim} \Delta=J+\nu\,,
\ee where $\nu=\sum_{n}N_n$ and $J$ is the R-charge emerging
from $SO(10)\rightarrow SO(8)$ and $N_n$ is the number of string excitations. Even though, the Penrose limit requires {\it inter alia}
going to large radius so that the resulting BMN formula
(\ref{bmndim}) is expected to be only valid for states with large
R-charge $J$, (\ref{bmndim}) can be extrapolate to $\lambda
\approx 0$ for all $J$'s.

\section{N=4 SYM spectrum: Polya(kov) Theory}

In order to make a comparison of our  previous results with the
${\cal N}=4$ SYM spectrum we have to devise strategy to enumerate
SYM states, and the correct way to proceed is to use Polya theory
\cite{Polya}. The idea was first applied by A. Polyakov in
\cite{Polyakov:2001af} to the counting of gauge invariant
operators made out only of bosonic "letters".

Let us start by briefly reviewing the basics of Polya theory.
Consider a set of words $A, B,\dots$ made out of $n$ letters
chosen within the alphabet $\{a_{i}\}$ with $i=1,\dots p$. Let $G$
be a group action defining the equivalence relation $A\sim B$ for
$A=gB$ with $g\in G\subset S_{n}$. Elements $g\in S_{n}$ can be
divided into conjugacy classes $[g]=(1)^{b_1}\dots (n)^{b_n}$,
according to the numbers $\{b_{k}(g)\}$ of cycles of length $k$.
Polya theorem states that the set of inequivalent words are
generated by the formula:
\be\label{polya}
P_{G}(\{a_i\})\equiv\f{1}{|G|}\sum_{g\in G}
\prod_{k=1}^{n}(a_1^k+a_2^k+\dots +a_p^k)^{b_{k}(g)}\,.
\ee

In particular, for $G=Z_{n}$, the cyclic permutation subgroup of
$S_{n}$, the elements $g\in G$ belong to one of the conjugacy
classes $[g]=(d)^{\f{n}{d}}$ for each divisor $d$ of $n$. The
number of elements in a given conjugacy class labelled by $d$ is
given by Euler's totient function $\varphi(d)$, that equals the
number of integers relatively prime to and smaller than $n$.  For
$n=1$ one defines $\varphi(1)=1$. Computing $P_{G}$ for $G=Z_{n}$
one finds: \be P_{n}(\{a_i\})\equiv\f{1}{n}\sum_{d/n}\varphi(d)
(a_1^d+a_2^d+\dots +a_p^d)^{\f{n}{d}}\,.\overline{} \ee

The number of inequivalent words can be read off from
(\ref{polya}) by simply letting $a_{i}\rightarrow 1$.

For instance, the possible choices of "necklaces" with six "beads"
of two different colors $a$ and $b$, are given by
\[
P_{6}(a,b)=\f1 6[(a+b)^6+(a^2+b^2)^3+2(a^3+b^3)^3+2(a^6+b^6)]=
\]
\[
=a^6+a^5b+3a^4b^2+4a^3b^3+3a^2b^4+ab^5+b^6\,,
\]
and the number of different necklaces is $P_{6}(a=b=1)=14$.

We are now ready to implement this construction in ${\cal N}=4$
theory, where the letters are the fundamental  fields together
with their derivatives
\[
\d^s\varphi^i; \d^s\lambda^A; \d^s\bar{\lambda}_A; \d^sF^+;
\d^sF^-\,.
\]

There are the ${\bf 6}$ scalar fields $\varphi^i$, ${\bf 4}$ Weyl
gaugini $\lambda^A$, ${\bf 4}^*$ conjugate ones $\bar{\lambda}_A$
and the (anti) self-dual field strengths $F^\pm$. Since it is
irrelevant for counting operators wether one is in free theory or
not, we take as mass-shell conditions the free field
equations. The single-letter on-shell partition functions then
take the following form \be Z_{s}(q)=\sum_{\Delta}n_{\Delta}^s
q^{\Delta}=n_{s} \f{q^{\Delta_{s}}}{(1-q)^4}(1-q^2)\,,\qquad
\Delta_s=1\,, \ee for the scalars such that $\d^\mu\d_\mu \varphi=0$. \be
Z_f(q)=\sum_{\Delta}n^f_{\Delta}q^{\Delta}=n_f\f{2q^{\Delta_f}}
{(1-q)^4}(1-q)\,,\qquad \Delta_f=3/2\,, \ee for the fermions with
$\gamma^\mu\d_\mu\lambda =0$. For the vector field one gets a little bit
involved expression, because apart from the equation of motion one
has to take into account Bianchi identities  $\d^\mu F^+_{\mu\nu}=\d^\mu F^-_{\mu\nu}=0$ \be
Z_v(q)=\sum_{\Delta}n_{\Delta}^v q^{\Delta}= n_v \f{2q^{\Delta_f}}
{(1-q)^4}(3-4q+q^2)\,,\qquad \Delta_f=2\,. \ee

Taking statistics into account for $U(N)$  we obtain the free SYM
partition function \be\label{Zym}
Z_{YM}=\sum_{M=1}^{N}\sum_{d|n}\f{1}{n}\varphi(d)
[Z_{s}(q^d)++Z_v(q^d)-Z_f(q^d)-Z_{\bar{f}}(q^d)]^{n/d}\,. \ee In
fact, (\ref{Zym}) is not exactly a partition function, rather it is
what one may call the Witten index, wherein fermions enter with a
minus sign and bosons with a plus sign. Now, words that consist of
more than $N$ constituents decompose into multi-trace operators,
so representing $n=kd$, where $d$ is a divisor of $n$ and summing
over $k$ and $d$ independently in the limit $N\to\infty$ one gets
\be Z_{YM}=-\sum_{d}\f{\varphi(d)}{d}\log
[1-Z_s(q^d)-Z_v(q^d)+Z_f(q^d)+Z_{\bar{f}}(q^d)]\,. \ee

For $SU(N)$, we have to subtract words that consist of  a single
constituent, which are not gauge invariant, thus the sum starts
with $n=2$ or equivalently \be Z^{SU(N)}=Z^{U(N)}-Z^{U(1)}\,. \ee

Finally, for ${\cal N}=4$ we have $n_s=6, n_f=n_{\bar{f}}=4, n_v=1$. Plugging
into (\ref{Zym}) and expanding in powers of $q$ up to $\Delta=4$ yields
\be
Z_{{\cal N}=4}(q)=21q^2-96q^{5/2}+361q^3-1328q^{7/2}+4601q^4+\dots
\ee

\subsection{Eratostene's (super)sieve}

In order to simplify the comparison of the spectrum of
SYM with the previously derived string spectrum, one can restrict
the attention to superconformal primaries by means of
Eratostenes's super-sieve,  that allows us to get rid of the
superdescendants. This procedure would be trivial if we knew that
all multiplets were long, but unfortunately the partition function
contains $1/2$-BPS, $1/4$-BPS, semishort ones too. The
structure of these multiplets of $PSU(2,2|4)$ is more elaborated
than the structure of long multiplets, which in turn is simply
coded in and factorizes on the highest weight state.

Superconformal primaries, \ie HWS of $SU(2,2|4)$, are
defined by the condition
\be
\hat{\delta}_{S}{\cal O}\equiv [\xi_{A}S^A+\bar{\xi}^A\bar{S}_A,{\cal O}]=0\,,
\ee
where $\delta$-is the supersymmetry transformation
\[
\hat{\delta}_S=\delta_S-\delta_Q\,,\qquad (\eta=x-\xi)
\]
\[
\hat{\delta}_S\varphi^i=0\,,\quad \hat{\delta}_S\lambda^A=
\tau_i^{AB}\varphi^i\xi_B\,,\quad \hat{\delta}_S\bar{\lambda}_A=0
\,,\quad \hat{\delta}_S F_{\mu\nu}=\xi_A\sigma_{\mu\nu}\lambda^A
\]
and $\tau_i^{AB}$ are $4\times 4$ Weyl blocks of Dirac matrices in
$d=6$. The procedure can be implemented step by step using computer.
\begin{itemize}
\item Start with the lowest primaries -- the Konishi
scalar field $K_1=tr\varphi_{i}\varphi^i$, and the lowest CPO
$Q_{20'}^{ij}=tr\varphi^{(i}\varphi^{j)|}$)
\item Remove their superdescendants
\item The first operator one finds at the lowest dimension is
necessarily a superprimary.
\item Go back to step 2
\end{itemize}

We have been able to perform this procedure up to $\Delta=11.5$
and the agreement with "naive" superstring spectrum
\[
H_{l}=\sum_{n,l}[0n0]_{(00)}\times \hat{H}_l^{flat}
\]
is perfect! Let us stress once more that  our mass formula
(\ref{bmndim}), though derived exploing HS symmetry and suggested
by the BMN formula extrapolated to the HS enhancement point,
reproduces semishort as well as genuinely long multiplets. The
latter correspond to massive string states which never get close
to being massless.

\section{HS extension of (P)SU(2,2$|$4)}

In the second part of this lecture, we identify the HS content of
${\cal N}=4$ SYM at the HS enhancement point.  Since we focus on
the higher spin extension of superconformal algebra, it is
convenient to realize $SU(2,2|4)$ by means of (super)oscillators
$\zeta_{\Lambda}= (y_{a},\theta_A)$ with \be
[y_a,\bar{y}^b]=\delta_{a}{}^{b}\,,\qquad
\{\theta_A,\bar{\theta}^B\}=\delta_{A}{}^{B}\,, \ee where $y_a,
\bar{y}^b$ are bosonic oscillators with $a,b=1,\dots, 4$ and
$\theta_A, \bar{\theta}^B$ are fermionic oscillators with
$A,B=1,\dots, 4$. The $su(2,2|4)$ superalgebra is spanned by
various traceless bilinears of these oscillators. There are
generators, \be J_{a}{}^{b}=\bar{y}^a
y_{b}-\f{1}{2}K\delta^{a}{}_{b}\,,\qquad K=\f{1}{2}\bar{y}^a y_a
\ee which represents $so(4,2)\oplus u(1)_K$ subalgebra and
generators \be
T^{A}{}_{B}=\bar{\theta}^A\theta_B-\f{1}{2}B\delta^{A}{}_{B}\,,\qquad
B=\f{1}{2}\bar{\theta}^A\theta_A \ee which correspond $su(4)\oplus
u(1)_B$. The abelian charge $B$ is to be identified with the
generator of Intriligator's "bonus symmetry" dual to the anomalous
$U(1)_B$ chiral symmetry of type IIB in the $AdS$ bulk. The
Poincar\'e and superconformal supercharges are of the form \be
Q_{a}^A=y_a\theta^A\,,\qquad \bar{Q}^a_A=\bar{y}^a\theta_A\,. \ee
The combination \be
C=K+B=\f{1}{2}\bar{\zeta}^{\Lambda}\zeta_{\Lambda}\,. \ee is a
central charge that commutes with all the other generators. Since
all of the fundamental fields $\{A_{\mu}, \lambda_{A}^{\alpha},
\bar{\lambda}^A_{\dot{\alpha}}, \varphi^i\}$ have central charge
equal to zero, we expect that local composites, just as well, have
central charge equal to zero. So we consistently put \be C=0\,.
\ee

The higher spin extension $hs(2,2|4)$ is generated by the odd
powers of the above generators
\be
hs(2,2|4)=Env(su(2,2|4))/I_C=\bigoplus_{l=0}^{\infty} A_{2l+1}\,,
\ee
where $I_C$ is the ideal generated by $C$ and the elements
$J_{2l+1}$ in $A_{2l+1}$ are of the form
\be
J_{2l+1}=P^{\Lambda_1\dots \Lambda_{2l+1}}_{\Sigma_1\dots
\Sigma_{2l+1}} \bar{\zeta}^{\Sigma_1}\dots
\bar{\zeta}^{\Sigma_{2l+1}}
\zeta_{\Lambda_1}\dots\zeta_{\Lambda_{2l+1}}-\textnormal{traces}\,.
\ee

The singleton representation of $su(2,2|4)$  turns out to be also
the singleton of $hs(2,2|4)$ in such a way that any state in the
singleton representation of $hs(2,2|4)$ can be reached from the
HWS by one step using a single higher spin generator. Note, that
in $su(2,2|4)$ the situation is different, namely in order to
reach a generic descendant from the HWS one has to apply several
times different rasing operators.

\subsection{sl(2) and its HS extension hs(1,1)}

Since the $hs(2,2|4)$ algebra is rather  complicated, in order to
clarify the algebraic construction, we make a short detour in what
may be called the $hs(1,1)$ algebra, the higher spin extension of
$sl(2)\approx su(1,1)$.

Consider the $sl(2)$ subalgebra: \be [J_-, J_+]=2J_3\,,\qquad
[J_3,J_{\pm}]=\pm J_{\pm}\,. \ee This algebra can be represented
in terms of oscillators \be J_+=a^++a^+a^+a\,,\qquad
J_3=\f{1}{2}+a^+a\,,\qquad J_-=a\,, \ee where, as usual,
$[a,a^+]=1$ and the vacuum state $\vac$  is annihilated by
$J_-=a$. Other $sl(2)$ HWS's are defined by \be
J_-f(a^+)\vac=0\quad\Rightarrow\quad f(a^+)=1\,. \ee Any state
$(a^+)^n\vac$ in this defining representation can be generated
from its HWS $\vac$ by acting with $J_+^n$. Therefore $f(a^+)$
defines a single irreducible representation of $sl(2)$, which will
be called {\it{singleton}} and denoted by $V_F $. The $sl(2)$ spin
of $V_F$ is $-J_3\vac=-\f{1}{2}\vac$. The dynamics of this
subsector is governed by a Heisenberg spin chain.

The embedding of $sl(2)$ in ${\cal N}=4$ SYM can be performed in
different ways. In particular, the HWS can be identified with the
scalar $Z=\varphi^5+i\varphi^6$ and its $sl(2)$ descendants can be
generated by the action of the derivative along a chosen complex
direction, for instance $D=D_1+iD_2$,
\be
(a^+)^n\vac\quad\leftrightarrow\quad D^nZ\,.
\ee
The tensor product of $L$ singletons may be represented in the
space of functions $f(a_{(1)}^+,\dots, a_{(L)}^+)$. The resulting
representation is no longer irreducible. This can be seen by
looking for $sl(2)$ HWS's
\be
J_-f(a_{(1)}^+,\dots, a_{(L)}^+)=\sum_{s=1}^{L}\d_s
f(a_{(1)}^+,\dots, a_{(L)}^+)=0\,.
\ee
There is indeed more than one solution to these equations given
by all possible functions of the form $f_L(a_{(s)}^+-a_{(s')}^+)$.
The basis for $sl(2)$ HWS's can be taken to be
\be
|j_1,\dots, j_{L-1}\rangle=(a_{(L)}^+-a_{(1)}^+)^{j_1}\dots
(a_{(L)}^+-a_{(L-1)}^+)^{j_{L-1}}\vac\,,
\ee
with spin $J_3=\f{1}{2}+\sum_s j_s$. In particular for $L=2$ one
finds the known result
\be
V_F\times V_F=\sum_{j=0}^{\infty}V_j\,,
\ee
where $V_j$ is generated by acting with $J_+$ on the HWS
$|j\rangle=(a_{(2)}^+-a_{(1)}^+)^j\vac$.

The higher spin algebra $hs(1,1)$ is generated by operators of the form
\be
J_{p,q}=(a^+)^p a^q + ... \,.
\ee
The generators $J_{p,q}$ with $p<q$ are raising operators. In the
tensor product of $L$ singletons, HWS's of $hs(1,1)$ are the
solutions of
\be
\sum_{i=1}^L (a_{(i)}^+)^p \d_{i}^q f(a_{(1)}^+,\dots,
a_{(L)}^+)=0\,,\qquad p<q\,.
\ee
For $L=2$ we can easily see that only two out of this infinite tower of
HWS's survive for $j=0$ and $j=1$. That is all even
objects belong to the same higher spin multiplet and all odd ones
belong to another multiplet. For $L>2$ one may consider either
totally symmetric or totally antisymmetric representations. It can
be easily shown that all of them are HWS's of HS multiplet.
\be
\begin{picture}(22,12)(-1,1)
{\linethickness{0.210mm}
\put(0,10){\line(1,0){40}} 
\put(00,00){\line(1,0){40}} 
\put(00,00){\line(0,1){10}} 
\put(10,00){\line(0,1){10}} 
\put(20,00){\line(0,1){10}}
\put(30,00){\line(0,1){10}}
\put(40,00){\line(0,1){10}} }
\end{picture}\qquad\Rightarrow \vac_{(L)}\sim Z^L\,,
\begin{picture}(12,32)(-1,15)
{\linethickness{0.210mm}
\put(0,0){\line(0,1){40}}
\put(10,0){\line(0,1){40}}
\put(0,0){\line(1,0){10}}
\put(0,40){\line(1,0){10}}
\put(0,10){\line(1,0){10}}
\put(0,20){\line(1,0){10}}
\put(0,30){\line(1,0){10}}
}
\end{picture}
: \prod_{i<j}(a_i^+-a_j^+)\vac_{(L)}\sim (ZDZ\dots
D^{L-1}Z+a.s.)\,. \ee For more complicated Young tableaux,  where
$L$ boxes distributed in different $k$ columns it can be shown,
that there is a solution of the form \be\label{YT}
\begin{picture}(50,50)(-1,15)
{\linethickness{0.210mm}
\put(0,0){\line(0,1){50}}
\put(0,50){\line(1,0){50}}
\put(0,40){\line(1,0){50}}
\put(0,30){\line(1,0){30}}
\put(0,20){\line(1,0){20}}
\put(10,0){\line(0,1){50}}
\put(20,20){\line(0,1){30}}
\put(30,30){\line(0,1){20}}
\put(50,40){\line(0,1){10}}
\put(0,0){\line(1,0){10}}
}
\end{picture}
=\prod_{p=1}^{k}\prod_{i_p<j_p}^{L_p}(a_{i_p}^+-a_{j_p}^+)\vac_{(L)}
\leftrightarrow Z^{n_1} (DZ)^{n_2}\dots
(D^{n_s}Z)+\textnormal{perms}\,. \ee The fact that (\ref{YT}) is
indeed a  solution is easily derived, however, its uniqueness is
hard to prove.

The generalization to $hs(2,2|4)$ is almost straightforward for
the totally symmetric representation
\[
\begin{picture}(52,12)(-1,0)
{\linethickness{0.210mm}
\put(0,10){\line(1,0){40}} 
\put(00,00){\line(1,0){40}} 
\put(00,00){\line(0,1){10}} 
\put(10,00){\line(0,1){10}} 
\put(20,00){\line(0,1){10}}
\put(30,00){\line(0,1){10}}
\put(40,00){\line(0,1){10}} }
\end{picture}\textnormal{HWS:}\quad \vac_{(L)}\leftrightarrow
Z^L\,.
\]
Namely, one starts with 1-impurity states \be
(WZ^{L-1}+\textnormal{symm.})=\f{1}{L}J_{W\bar{Z}}^{HS}Z^L \,, \ee
where the impurity $(W)$  appears symmetrically in all places, and
proceeds with 2-impurity states \be (W_1
Z^{k-2}W_2Z^{L-k}+\textnormal{symm.})=\f{1}{L(L-1)}J_{W_1\bar{Z}}^{HS}
J_{W_2\bar{Z}}^{HS}Z^L \,, \ee and so on. Note, that all operators
of  this symmetry are descendants of $Z^L$ due to the fact, that
each state in a singleton representation can be reached by a
single step starting from the highest weight state.

For generic Young tableaux the task is more involved.  However,
the above construction goes through and the same arguments hold.
For example, besides the descendants $J_{W\bar{Z}}$ of $Z^L$ there
are $L-1$ 1-impurity multiplets of states associated to the $L-1$
Young tableaux with $L-1$ boxes in the first row and a single box
in the second one. The vacuum state of HS multiplets associated to
such tableaux can be taken to be $Y_{(k)}\equiv Z^k Y
Z^{L-k-1}-YZ^{L-1}$ with $k=1,\dots , L-1$. Any state with one
impurity $Z^k W Z^{L-k-1}-WZ^{L-1}$ can be found by acting on
$Y_{(k)}$ with the HS generators $J_{W\bar{Y}}$, where
$J_{W\bar{Y}}$ is the HS generator that transforms $Y$ into $W$
and annihilates $Z$. The extension to other Young tableaux
proceeds similarly though tediously.

\subsection{HS content of ${\cal N}=4$ SYM $\sim$ IIB  at HS enhancement point}

The free ${\cal N}=4$ singleton partition function  is given by
the expression \be Z\begin{picture}(10,10)(-1,4)
\put(0,0){\line(0,1){5}} \put(0,5){\line(1,0){5}}
\put(0,0){\line(1,0){5}} \put(5,0){\line(0,1){5}}
\end{picture}
=\sum_{\Delta_s}(-1)^{2s}d_{\Delta_s}q^{\Delta_s}=
\f{2q(3+\sqrt{q})}{(1+\sqrt{q})^3}\,, \ee where $\Delta_s$ is the
bare conformal dimension, \ie  the dimension at the HS enhancement
point. Note that the singleton is not gauge invariant, thus one
should build gauge invariant composites with two or more
"letters". For example, the symmetric doubleton
\begin{picture}(16,10)(0,0)
\put(0,0){\line(0,1){7}}
\put(0,7){\line(1,0){14}}
\put(0,0){\line(1,0){14}}
\put(7,0){\line(0,1){7}}
\put(14,0){\line(0,1){7}}
\end{picture}
can be obtained multiplying two singletons
\be
\begin{picture}(10,10)(0,1)
\put(0,0){\line(0,1){10}}
\put(0,10){\line(1,0){10}}
\put(0,0){\line(1,0){10}}
\put(10,0){\line(0,1){10}}
\end{picture}
\times
\begin{picture}(10,10)(1,1)
\put(0,0){\line(0,1){10}}
\put(0,10){\line(1,0){10}}
\put(0,0){\line(1,0){10}}
\put(10,0){\line(0,1){10}}
\end{picture}
=
\begin{picture}(20,10)(1,1)
\put(0,0){\line(0,1){10}}
\put(0,10){\line(1,0){20}}
\put(0,0){\line(1,0){20}}
\put(10,0){\line(0,1){10}}
\put(20,0){\line(0,1){10}}
\end{picture}
+
\begin{picture}(20,10)(1,7)
\put(0,0){\line(0,1){20}}
\put(0,10){\line(1,0){10}}
\put(0,0){\line(1,0){10}}
\put(10,0){\line(0,1){20}}
\put(0,20){\line(1,0){10}}
\end{picture}
\,, \ee where the antisymmetric diagram appears only in
interactions and must be neglected in the free theory. The
spectrum of single-trace operators in ${\cal N}=4$ SYM theory with
$SU(N)$ gauge group  is given by all possible cyclic words built
from letters chosen from $Z$
\begin{picture}(5,10)(2,2)
\put(0,0){\line(0,1){5}}
\put(0,5){\line(1,0){5}}
\put(0,0){\line(1,0){5}}
\put(5,0){\line(0,1){5}}
\end{picture}. It can be computed using Polya  theory \cite{Polya}, which gives
the generating function
\be\label{partfunc}
Z(q,y_i)=\sum_{n>2} Z_n (q,y_i)=\sum_{n>2,d|n}
u^n\f{\varphi(d)}{n} Z
\begin{picture}(10,10)(-1,4)
\put(0,0){\line(0,1){5}}
\put(0,5){\line(1,0){5}}
\put(0,0){\line(1,0){5}}
\put(5,0){\line(0,1){5}}
\end{picture}
(q^d, y_i^d)^{\f{n}{d}} \ee for cyclic words. The sum runs over
all integers $n>2$ and their divisors $d$, and $\varphi(d)$ is
Euler's  totient function, defined previously. The partition
function (\ref{partfunc}) can be decomposed in representations of
$hs(2,2|4)$, \ie HS multiplets. In particular, all operators
consisting of two letters, assemble into the (symmetric)
doubleton. \be Z_2\equiv Z^{\delta_{a,b}}
\begin{picture}(0,10)(15,4)
\put(0,0){\line(0,1){5}}
\put(0,5){\line(1,0){10}}
\put(0,0){\line(1,0){10}}
\put(5,0){\line(0,1){5}}
\put(10,0){\line(0,1){5}}
\end{picture}
=\sum_{n}\chi(v_{2n})\,. \ee For tri-pletons with three letters,
one finds the  totally symmetric tableau and the totally
antisymmetric one. \be Z_{3}=Z^{(d_{abc})}
\begin{picture}(0,10)(22,4)
\put(0,0){\line(0,1){5}}
\put(0,5){\line(1,0){15}}
\put(0,0){\line(1,0){15}}
\put(5,0){\line(0,1){5}}
\put(10,0){\line(0,1){5}}
\put(15,0){\line(0,1){5}}
\end{picture}
+Z^{(f_{abc})}
\begin{picture}(0,10)(20,15)
\put(0,0){\line(0,1){15}}
\put(0,5){\line(1,0){5}}
\put(0,0){\line(1,0){5}}
\put(0,15){\line(1,0){5}}
\put(5,0){\line(0,1){15}}
\put(0,10){\line(1,0){5}}
\end{picture}
=\sum_{k=-\infty}^{+\infty} \sum_{n=0}^{\infty} c_n\Big(
\chi(v_{2k}^{n})+\chi(v_{2k+1}^{n+3})+\chi(v_{2k+1}^{n})+
\chi(v_{2k}^{n+3})\Big)\,, \ee where the coefficients $c_n\equiv
1+[n/6]-\delta_{n,1}$, yielding the multiplicities of $psu(2,2|4)$
multiplets inside $hs(2,2|4)$, count the number of ways one can
distribute derivatives (HS descendants) among the boxes of the
tableaux.  Similarly for the tetra-pletons and penta-pletons one
finds

\be
Z_{4}=Z^{(q_{abcd})}
\begin{picture}(0,10)(25,4)
\put(0,0){\line(0,1){5}}
\put(0,5){\line(1,0){20}}
\put(0,0){\line(1,0){20}}
\put(5,0){\line(0,1){5}}
\put(10,0){\line(0,1){5}}
\put(15,0){\line(0,1){5}}
\put(20,0){\line(0,1){5}}
\end{picture}
+Z^{(d\cdot f)}
\begin{picture}(0,10)(20,15)
\put(0,0){\line(0,1){15}}
\put(0,5){\line(1,0){5}}
\put(0,0){\line(1,0){5}}
\put(0,15){\line(1,0){10}}
\put(5,0){\line(0,1){15}}
\put(0,10){\line(1,0){10}}
\put(10,10){\line(0,1){5}}
\end{picture}
+Z^{(f\cdot f)}
\begin{picture}(0,10)(20,10)
\put(0,0){\line(0,1){10}}
\put(0,0){\line(1,0){10}}
\put(0,10){\line(1,0){10}}
\put(10,0){\line(0,1){10}}
\put(5,0){\line(0,1){10}}
\put(0,5){\line(1,0){10}}
\end{picture}
\ee
\be
Z_5=Z^{(p_{abcde})}
\begin{picture}(0,20)(30,4)
\put(0,0){\line(0,1){5}}
\put(0,5){\line(1,0){25}}
\put(0,0){\line(1,0){25}}
\put(5,0){\line(0,1){5}}
\put(10,0){\line(0,1){5}}
\put(15,0){\line(0,1){5}}
\put(20,0){\line(0,1){5}}
\put(25,0){\line(0,1){5}}
\end{picture}
+Z
\begin{picture}(15,10)(0,9)
\put(0,0){\line(0,1){10}}
\put(0,0){\line(1,0){10}}
\put(0,10){\line(1,0){15}}
\put(10,0){\line(0,1){10}}
\put(5,0){\line(0,1){10}}
\put(0,5){\line(1,0){15}}
\put(15,5){\line(0,1){5}}
\end{picture}
+2Z
\begin{picture}(15,10)(0,15)
\put(0,0){\line(0,1){15}}
\put(0,5){\line(1,0){5}}
\put(0,0){\line(1,0){5}}
\put(0,15){\line(1,0){15}}
\put(5,0){\line(0,1){15}}
\put(0,10){\line(1,0){15}}
\put(10,10){\line(0,1){5}}
\put(15,10){\line(0,1){5}}
\end{picture}
+Z
\begin{picture}(10,10)(0,10)
\put(0,-5){\line(0,1){15}}
\put(0,0){\line(1,0){10}}
\put(0,10){\line(1,0){10}}
\put(10,0){\line(0,1){10}}
\put(5,-5){\line(0,1){15}}
\put(0,5){\line(1,0){10}}
\put(0,-5){\line(1,0){5}}
\end{picture}
+Z
\begin{picture}(10,10)(0,25)
\put(0,0){\line(0,1){25}}
\put(5,0){\line(0,1){25}}
\put(0,0){\line(1,0){5}}
\put(0,5){\line(1,0){5}}
\put(0,10){\line(1,0){5}}
\put(0,15){\line(1,0){5}}
\put(0,20){\line(1,0){5}}
\put(0,25){\line(1,0){5}}
\end{picture}
\ee
\newline
In the above partition functions,  totally symmetric tableaux are
to be associated to KK descendants of the HS doubleton multiplet.
Other tableaux are those associated to lower spin St\"{u}ckelberg
multiplets, that are needed in order for {\it ''La Grande Bouffe"}
to take place. We checked that multiplicities, quantum numbers and
naive dimensions are correct so that they can pair with massless
multiplets and give long multiplets. Finally there are genuinely
massive representations that decompose into long multiplets of
$su(2,2|4)$ even at the HS point.

\section{Conclusions and outlook}

Let us summarize the results presented in the lecture.
\begin{itemize}
\item There is perfect agreement between the string spectrum on
$AdS_5\times S^5$ "extrapolated" to  the HS enhancement point with
the spectrum of single trace gauge invariant operator in free
${\cal N}=4$ SYM at large $N$. \item The massless doubleton
comprises the  HS gauge fields which are dual to the classically
conserved HS currents. Massive YT-pletons, \ie multiplets
associated to Young tableau compatible with gauge invariance,
correspond to KK excitations, St\"{u}ckelberg fields and genuinely
long and massive HS multiplet. {\it ''La Grande Bouffe"} is
kinematically allowed to take place at $\lambda\ne 0$. \item The
one loop anomalous dimensions of the HS currents are given by
\[
\gamma_S^{^{\rm 1-loop}}=\sum_{k=1}^S \f{1}{k}\,,
\]
and it looks likely that it have a number theoretical origin.
\item There are some interesting issues of integrability. First of
all the dilatation operator can be identified with the Hamiltonian
of a superspin chain and is integrable at one loop or in some
sectors up to two and three loops. Flat currents in $AdS_{5}\times
S^5$ give rise to a Yangian structure. Finally, the HS gauge
theory can be formulated as  a Cartan integrable system. \item
There are some surprising features in ${\cal N}=4$ SYM that have
emerged from resolving the operator mixing at finite $N$
\cite{BRS}. In particular there are ''unprotected" operators with
$\gamma_S^{^{\rm 1-loop}}=0$ and there are operators with
nonvanishing anomalous dimension, whose large $N$ expansion
truncates at some finite order in $N$ \cite{JFM}
\[
\gamma_s^{\textnormal{1-loop}}=a+\f{b}{N}+\f{c}{N^2}\,,
\]
with no higher order terms in $1/N$.
\end{itemize}

These and other facets of the AdS/CFT at small radius are worth
further study in connection with integrability and HS symmetry
enhancement. Sharpening the worldsheet description of the dynamics
of type IIB superstrings on $AdS_{5}\times S^5$ may turn to be
crucial in all the above respects. Twelve dimensional aspects and
two-time description \cite{Bars} are worth exploring, too.

\section*{Acknowledgements}
Most of the material in this lecture is based on  work done by M.
B. in collaboration with Niklas Beisert, Jos\'e Francisco Morales
Morera and Henning Samtleben. M.B.  would like to thank them as
well as Dan Freedman, Mike Green, Stefano Kovacs, Giancarlo Rossi,
Kostas Skenderis and Yassen Stanev for fruitful collaborations on
the holographic correspondence, superconformal gauge theories and
higher spins. Let us also acknowledge Misha Vasiliev, Per Sundell,
Ergin Sezgin, Augusto Sagnotti, Fabio Riccioni, Tassos Petkou and
Dario Francia for stimulating discussions on higher spins. Last
but not least, let us thank Glenn Barnich, Giulio Bonelli, Maxim
Grigoriev and Marc Henneaux, the organizers of the Solvay
Conference on ''Higher Spins", for creating a very stimulating
atmosphere and especially for their patience with the proceedings.
The work of M.B. was supported in part by INFN, by the MIUR-COFIN
contract 2003-023852, by the EU contracts MRTN-CT-2004-503369 and
MRTN-CT-2004-512194, by the INTAS contract 03-516346 and by the
NATO grant PST.CLG.978785. The work of V.D. was supported by
grants RFBR 02-02-17067, the Landau Scholarship Foundation,
Forschungszentrum J\"u\-lich and the Dynasty Scholarship
Foundation.


\begin{thebibliography}{10}
\ifx\href\asklfhas\newcommand{\href}[2]{#2}\fi \raggedright \small
\parskip 0pt


\bibitem{Bianchi:2003wx}
M.~Bianchi, J.~F.~Morales and H.~Samtleben,
\textsf{JHEP~0307,~062~(2003)},
\href{http://arXiv.org/abs/hep-th/0305052}{\texttt{hep-th/0305052}}.

\bibitem{Beisert:2003te}
N.~Beisert, M.~Bianchi, J.~F.~Morales and H.~Samtleben,
\textsf{JHEP~0402,~001~(2004)},
\href{http://arXiv.org/abs/hep-th/0310292}{\texttt{hep-th/0310292}}.

\bibitem{Beisert:2004di}
N.~Beisert, M.~Bianchi, J.~F.~Morales and H.~Samtleben,
\textsf{JHEP {\bf 0407}, 058 (2004)}
 \href{http://arXiv.org/abs/hep-th/0405057}{\texttt{hep-th/0405057}}.

\bibitem{MBstring}
  M.~Bianchi,
  Comptes Rendus Physique {\bf 5}, 1091 (2004)
  [arXiv:hep-th/0409292].

\bibitem{Aharony:1999ti}
O.~Aharony, S.~S.~Gubser, J.~M.~Maldacena, H.~Ooguri and Y.~Oz,
Phys.\ Rept.\  {\bf 323} (2000) 183 [arXiv:hep-th/9905111].

\bibitem{D'Hoker:2002aw}
E.~D'Hoker and D.~Z.~Freedman, ``Supersymmetric gauge theories and
the AdS/CFT correspondence,'' arXiv:hep-th/0201253.

\bibitem{Bianchi:2000vh}
M.~Bianchi,
Nucl.\ Phys.\ Proc.\ Suppl.\  {\bf 102}, 56 (2001)
[arXiv:hep-th/0103112].

\bibitem{Tseytlin:2003ii}
A.~A.~Tseytlin, ``Spinning strings and AdS/CFT duality,''
arXiv:hep-th/0311139.

\bibitem{WittenJHS60}
E.~Witten, \textit{``Spacetime Reconstruction''}, Talk at JHS 60
Conference, Caltech, 3-4 Nov 2001
\href{http://quark.caltech.edu/jhs60/witten/1.html}.

\bibitem{Sundborg:1999ue}
B.~Sundborg,
\textsf{Nucl.~Phys.~B573,~349~(2000)},
\href{http://arXiv.org/abs/hep-th/9908001}{\texttt{hep-th/9908001}}.

\bibitem{Polyakov:2001af}
A.~M.~Polyakov,
\textsf{Int.~J.~Mod.~Phys.~A17S1,~119~(2002)},
\href{http://arXiv.org/abs/hep-th/0110196}{\texttt{hep-th/0110196}}.

\bibitem{Sezgin:2001zs}
E.~Sezgin and P.~Sundell,
\textsf{JHEP~0109,~036~(2001)},
\href{http://arXiv.org/abs/hep-th/0105001}{\texttt{hep-th/0105001}}.

\bibitem{Sezgin:2001yf}
E.~Sezgin and P.~Sundell,
\textsf{JHEP~0109,~025~(2001)},
\href{http://arXiv.org/abs/hep-th/0107186}{\texttt{hep-th/0107186}}.

\bibitem{Sezgin:2002rt}
E.~Sezgin and P.~Sundell,
\textsf{Nucl.~Phys.~B644,~303~(2002)},
\href{http://arXiv.org/abs/hep-th/0205131}{\texttt{hep-th/0205131}}.

\bibitem{Polya}
G.~P{\'o}lya and R.~Read, \textit{``Combinatorial enumeration of
groups, graphs, and chemical
  compounds''},
Springer-Verlag (1987), New-York, P{\'o}lya's contribution
translated from the German by Dorothee Aeppli.

\bibitem{Bouatta:2004kk}
N.~Bouatta, G.~Compere and A.~Sagnotti,
``An introduction to free higher-spin fields,''
arXiv:hep-th/0409068.

\bibitem{Petkoubrux}
  A.~C.~Petkou,
  ``Holography, duality and higher-spin theories,''
  arXiv:hep-th/0410116.

\bibitem{Vasbrux}
M.~Vasiliev, contribution to these Proceedings, in preparation.

\bibitem{Sundbrux}
  A.~Sagnotti, E.~Sezgin and P.~Sundell,
  ``On higher spins with a strong Sp(2,R) condition,''
  arXiv:hep-th/0501156.


\bibitem{Hullbrux}
  D.~Francia and C.~M.~Hull,
  ``Higher-spin gauge fields and duality,''
  arXiv:hep-th/0501236.

\bibitem{Vasiliev:2004qz}
M.~A.~Vasiliev, \textit{``Higher spin gauge theories in various
dimensions''},
\href{http://arXiv.org/abs/hep-th/0401177}{\texttt{hep-th/0401177}}.

\bibitem{Sorokin:2004ie}
D.~Sorokin, \textit{``Introduction to the classical theory of
higher spins''}, arXiv:hep-th/0405069.

\bibitem{MBRTN}
  M.~Bianchi,
  ``Higher spins and stringy AdS(5) x S(5),''
  arXiv:hep-th/0409304.

\bibitem{nobel} 
  D.~J.~Gross and F.~Wilczek,
  Phys.\ Rev.\ Lett.\  {\bf 30} (1973) 1343.
  H.~D.~Politzer,
  Phys.\ Rev.\ Lett.\  {\bf 30}, 1346 (1973).



\bibitem{Altarelli} 
  G.~Altarelli and G.~Parisi,
  Nucl.\ Phys.\ B {\bf 126}, 298 (1977).

\bibitem{Grib} 
  V.~N.~Gribov and L.~N.~Lipatov,
  Yad.\ Fiz.\  {\bf 15} (1972) 781
  [Sov.\ J.\ Nucl.\ Phys.\  {\bf 15} (1972) 438].


\bibitem{Petronzio}
G.~Curci, W.~Furmanski and R.~Petronzio,
Nucl.\ Phys.\ B {\bf 175}, 27 (1980).
W.~Furmanski and R.~Petronzio,
Phys.\ Lett.\ B {\bf 97}, 437 (1980).

\bibitem{DobrevPetkova} V.K. Dobrev and V.B. Petkova,
\textsf{Lett. Math. Phys. {\bf 9} (1985) 287-298};
\textsf{Fortschr. d. Phys. {\bf 35} (1987)537-572};
\textsf{Phys. Lett. {\bf 162B} (1985) 127-132}.

\bibitem{Dolan:2002zh}
F.~A.~Dolan and H.~Osborn,
\textsf{Ann.~Phys.~307,~41~(2003)},
\href{http://arXiv.org/abs/hep-th/0209056}{\texttt{hep-th/0209056}}.

\bibitem{Heslop:2003xu}
P.~J.~Heslop and P.~S.~Howe,
\textsf{JHEP~0401,~058~(2004)},
\href{http://arXiv.org/abs/hep-th/0307210}{\texttt{hep-th/0307210}}.

\bibitem{Andrianopoli:1998ut}
L.~Andrianopoli and S.~Ferrara,
\textsf{Lett.~Math.~Phys.~48,~145~(1999)},
\href{http://arXiv.org/abs/hep-th/9812067}{\texttt{hep-th/9812067}}.


\bibitem{Seiberg}
S.~M.~Lee, S.~Minwalla, M.~Rangamani and N.~Seiberg,
Adv.\ Theor.\ Math.\ Phys.\  {\bf 2}, 697 (1998)
[arXiv:hep-th/9806074].
  L.~F.~Alday, J.~R.~David, E.~Gava and K.~S.~Narain,
  arXiv:hep-th/0502186.

\bibitem{dinst}
M.~Bianchi, M.~B.~Green, S.~Kovacs and G.~Rossi,
JHEP {\bf 9808}, 013 (1998) [arXiv:hep-th/9807033].
N.~Dorey, T.~J.~Hollowood, V.~V.~Khoze, M.~P.~Mattis and
S.~Vandoren,
Nucl.\ Phys.\ B {\bf 552} (1999) 88 [arXiv:hep-th/9901128].

\bibitem{extrem}
E.~D'Hoker, D.~Z.~Freedman, S.~D.~Mathur, A.~Matusis and
L.~Rastelli,
arXiv:hep-th/9908160.
M.~Bianchi and S.~Kovacs,
Phys.\ Lett.\ B {\bf 468}, 102 (1999) [arXiv:hep-th/9910016].
B.~Eden, P.~S.~Howe, C.~Schubert, E.~Sokatchev and P.~C.~West,
Phys.\ Lett.\ B {\bf 472}, 323 (2000) [arXiv:hep-th/9910150].
E.~D'Hoker, J.~Erdmenger, D.~Z.~Freedman and M.~Perez-Victoria,
Nucl.\ Phys.\ B {\bf 589}, 3 (2000) [arXiv:hep-th/0003218].
B.~U.~Eden, P.~S.~Howe, E.~Sokatchev and P.~C.~West,
Phys.\ Lett.\ B {\bf 494}, 141 (2000) [arXiv:hep-th/0004102].

\bibitem{partial}
  B.~Eden, A.~C.~Petkou, C.~Schubert and E.~Sokatchev,
  Nucl.\ Phys.\ B {\bf 607}, 191 (2001)
  [arXiv:hep-th/0009106].
  G.~Arutyunov, B.~Eden, A.~C.~Petkou and E.~Sokatchev,
  Nucl.\ Phys.\ B {\bf 620}, 380 (2002)
  [arXiv:hep-th/0103230].

\bibitem{maldawilson} 
  J.~M.~Maldacena,
  Phys.\ Rev.\ Lett.\  {\bf 80}, 4859 (1998)
  [arXiv:hep-th/9803002].
  S.~J.~Rey and J.~T.~Yee,
  Eur.\ Phys.\ J.\ C {\bf 22}, 379 (2001)
  [arXiv:hep-th/9803001].

\bibitem{rainbow} 
  J.~K.~Erickson, G.~W.~Semenoff and K.~Zarembo,
  Nucl.\ Phys.\ B {\bf 582} (2000) 155
  [arXiv:hep-th/0003055].
  N.~Drukker and D.~J.~Gross,
  J.\ Math.\ Phys.\  {\bf 42}, 2896 (2001)
  [arXiv:hep-th/0010274].

\bibitem{bgkwilson}
M.~Bianchi, M.~B.~Green and S.~Kovacs,
JHEP {\bf 0204}, 040 (2002) [arXiv:hep-th/0202003].

\bibitem{holocthe}
  D.~Z.~Freedman, S.~S.~Gubser, K.~Pilch and N.~P.~Warner,
  Adv.\ Theor.\ Math.\ Phys.\  {\bf 3}, 363 (1999)
  [arXiv:hep-th/9904017].
  D.~Anselmi, L.~Girardello, M.~Porrati and A.~Zaffaroni,
  Phys.\ Lett.\ B {\bf 481}, 346 (2000)
  [arXiv:hep-th/0002066].

\bibitem{hensken} 
  M.~Henningson and K.~Skenderis,
  JHEP {\bf 9807}, 023 (1998)
  [arXiv:hep-th/9806087].

\bibitem{Bianchi:2001de}
M.~Bianchi, D.~Z.~Freedman and K.~Skenderis,
JHEP {\bf 0108}, 041 (2001) [arXiv:hep-th/0105276].
\bibitem{Bianchi:2001kw}
M.~Bianchi, D.~Z.~Freedman and K.~Skenderis,
Nucl.\ Phys.\ B {\bf 631}, 159 (2002) [arXiv:hep-th/0112119].

\bibitem{holoren}
  J.~Kalkkinen, D.~Martelli and W.~Muck,
  JHEP {\bf 0104}, 036 (2001)
  [arXiv:hep-th/0103111].
  D.~Martelli and W.~Muck,
  Nucl.\ Phys.\ B {\bf 654}, 248 (2003)
  [arXiv:hep-th/0205061].
  I.~Papadimitriou and K.~Skenderis,
  arXiv:hep-th/0404176.



\bibitem{superglu} 
  D.~Z.~Freedman, S.~S.~Gubser, K.~Pilch and N.~P.~Warner,
  JHEP {\bf 0007}, 038 (2000)
  [arXiv:hep-th/9906194].
  L.~Girardello, M.~Petrini, M.~Porrati and A.~Zaffaroni,
  Nucl.\ Phys.\ B {\bf 569} (2000) 451
  [arXiv:hep-th/9909047].
  M.~Bianchi, O.~DeWolfe, D.~Z.~Freedman and K.~Pilch,
  JHEP {\bf 0101}, 021 (2001)
  [arXiv:hep-th/0009156].


\bibitem{3point} 
  M.~Bianchi and A.~Marchetti,
  Nucl.\ Phys.\ B {\bf 686} (2004) 261
[arXiv:hep-th/0302019].
  M.~Bianchi, M.~Prisco and W.~Muck,
  JHEP {\bf 0311}, 052 (2003)
  [arXiv:hep-th/0310129].
  W.~Muck and M.~Prisco,
  JHEP {\bf 0404}, 037 (2004)
  [arXiv:hep-th/0402068].
  I.~Papadimitriou and K.~Skenderis,
  JHEP {\bf 0410}, 075 (2004)
  [arXiv:hep-th/0407071].

\bibitem{Berenstein:2002jq}
D.~Berenstein, J.~M.~Maldacena and H.~Nastase,
\textsf{JHEP~0204,~013~(2002)},
\href{http://arXiv.org/abs/hep-th/0202021}{\texttt{hep-th/0202021}}.

\bibitem{Berkovits:2002zk}
N.~Berkovits, ``ICTP lectures on covariant quantization of the
superstring,'' arXiv:hep-th/0209059.

\bibitem{Vas}
S.~E.~Konstein, M.~A.~Vasiliev and V.~N.~Zaikin,
\textit{``Conformal higher spin currents in any dimension and
AdS/CFT correspondence''}, \textsf{JHEP~0012,~018~(2000)},
\href{http://arXiv.org/abs/hep-th/0010239}{\texttt{hep-th/0010239}}.

\bibitem{Bianchi:2001cm}
  M.~Bianchi, S.~Kovacs, G.~Rossi and Y.~S.~Stanev,
  JHEP {\bf 9908}, 020 (1999)
  [arXiv:hep-th/9906188].
M.~Bianchi, S.~Kovacs, G.~Rossi and Y.~S.~Stanev,
\textsf{JHEP~0105,~042~(2001)},
\href{http://arXiv.org/abs/hep-th/0104016}{\texttt{hep-th/0104016}}.

\bibitem{Bianchi2loop}
M.~Bianchi, S.~Kovacs, G.~Rossi and Y.~S.~Stanev,
\textsf{Nucl.~Phys.~B584,~216~(2000)},
\href{http://arXiv.org/abs/hep-th/0003203}{\texttt{hep-th/0003203}}.

\bibitem{Beisert:2004ry}
  J.~A.~Minahan and K.~Zarembo,
  JHEP {\bf 0303}, 013 (2003)
  [arXiv:hep-th/0212208].
N.~Beisert, \textit{``The dilatation operator of N = 4 super
Yang-Mills theory and integrability''}, arXiv:hep-th/0407277.


\bibitem{BRS}
  M.~Bianchi, G.~Rossi and Y.~S.~Stanev,
  Nucl.\ Phys.\ B {\bf 685}, 65 (2004)
  [arXiv:hep-th/0312228].
  M.~Bianchi, B.~Eden, G.~Rossi and Y.~S.~Stanev,
  Nucl.\ Phys.\ B {\bf 646}, 69 (2002)
  [arXiv:hep-th/0205321].

\bibitem{JFM}
  S.~Bellucci, P.~Y.~Casteill, J.~F.~Morales and C.~Sochichiu,
  Nucl.\ Phys.\ B {\bf 707}, 303 (2005)
  [arXiv:hep-th/0409086].
  S.~Bellucci, P.~Y.~Casteill, J.~F.~Morales and C.~Sochichiu,
  ``Chaining spins from (super)Yang-Mills,''
  arXiv:hep-th/0408102.

\bibitem{Bars}
I.~Bars,
``Twistor superstring in 2T-physics,''
arXiv:hep-th/0407239.

\end{thebibliography}
\end{document}